\documentstyle[12pt]{article}
\begin{document}
\begin{center}
{\large \bf Vacuum fluctuations of a massless spin- $ 1\over 2 $ field
 around  multiple cosmic strings} \\[15mm]
{ \large A.N. Aliev $^{*}$, M.Horta\c csu $^{* \dagger }$ and
 N. \" Ozdemir $^{\dagger}$} \\[2mm]
$^{*}$  T\" UBITAK, Research Institute for Basic Sciences,\\
41470 Gebze, Turkey \\
$^{\dagger}$ ITU, Faculty of Sciences and Letters, Department of Physics,\\
 80626 Maslak, Istanbul, Turkey
\end{center}
\noindent
\vspace{1cm}

   We study the interaction of a massless quantized spinor field with
the gravitational field of $ N $ parallel static cosmic strings 
by using a perturbative approach. We show that the presence of more 
than one cosmic string gives rise to an additional contribution to 
the energy density of vacuum fluctuations, thereby leading to a 
vacuum force of attraction between two parallel cosmic strings.

\pagebreak

\section{Introduction}
    
    Cosmic strings predicted in the framework of various gauge theories
with spontaneously broken symmetries could have been created at
cosmological phase transitions in the early universe \cite{kib,vilsh}.
Since cosmic strings are produced at very large energy scales, one might
expect a highly curved spacetime around them. However in the case 
of a static and straight-line  cosmic string there exists a very simple 
exact solution of the Einstein field equations, that describes a locally 
flat conical spacetime around the string \cite{got}.  Furthermore, it has 
been shown that one may also construct an appropriate 
exact solution of the Einstein equations for a snapping cosmic string,
which serves as a source for spherical impulsive gravitational waves 
\cite{nutp}.

   Although the locally flat structure of the spacetime implies that
straight-line cosmic strings will not exert any local gravitational force
on surrounding particles, the particles ``interact'' with the global 
conical structure. It gives rise to the distinctive gravitational effects, 
such as lensing of distant objects, conical bremsstrahlung, etc. 
\cite{vil,ali}. On the other hand, it is well known that the non-trivial
topological structures restrict the modes of quantized fields propagating 
in locally flat spacetimes, thereby providing the appearance of vacuum
boundary effects \cite{birdav}.
One may, therefore, consider the conical spacetime around static and
straight-line cosmic strings as an attractive model for investigation  
the influence of gravitational field on the behaviour of quantized fields. 
In particular, such investigations for a massless quantized fields 
of different spins, propagating in the spacetime of a static 
cosmic string have been carried out in [8-10]. The propagation of a 
massless quantized scalar field in the spacetime of a snapping cosmic 
string was studied in \cite{hortac}.

   Recently in \cite{ali3} we calculated the effects of vacuum fluctuations
of a massless quantized electromagnetic field propagating in the spacetime of
multiple cosmic strings represented by  $ N $ parallel static (fixed),
straight-line strings. It has been shown that the presence of
more than one cosmic string provides an additional contribution to the
energy density of vacuum fluctuations, which results in the Casimir-like 
force of attraction between two parallel cosmic strings.

   The aim of the present paper is to extend the calculations of \cite{ali3}
to the case of a massless quantized spinor field. First, for an 
instructive purpose, we shall rederive the expression for the vacuum 
expectation value of the energy-momentum tensor of a massless 
spin-$ 1\over 2 $  field around a single cosmic string. 
For the case of multiple cosmic strings we shall adopt a perturbative 
approach, in which the gravitational field of the strings is treated 
as small metric perturbations about the Minkowskian spacetime. 
We construct, at first-order in metric perturbations, the Feynman 
propagator for the spinor field and evaluate the energy density 
arising from vacuum fluctuations. We shall also show that
at second-order perturbations of the metric, the effect of vacuum
polarization of the spinor field gives rise to a force of attraction 
between two parallel static and straight-line  cosmic strings.
Throughout the paper we use geometrical units, in which $ G=\,c=\,1$ and
$\hbar\approx 2.612\times{10}^{-66} cm^{2}$.

\section{ A single cosmic string }

The metric of a static and straight-line cosmic string lying along the z-axis of 
the cylindrical coordinate system is given by the interval
\begin{equation}
ds^2 = dt^2 - dz^2 - dr^2 - b^2\, r^2 \,d\theta^2
\label{met1}
\end{equation}
where $ b=1-4\mu $, and  $ \mu $ is the linear mass density of the string.
This metric is locally Minkowskian that is readily seen by passing
to a new angular coordinate $ \varphi \rightarrow  b\,\theta $. We have
\begin{equation}
ds^2 = dt^2 - dz^2 - dr^2 - r^2 \,d\varphi^2
\label{fmet1}
\end{equation}
However the global structure of this metric is a conical, 
as for the new azimuthal angle $ \varphi $ we have the following range

$$ 0\,\leq\,\varphi\,<\,2\pi(1-4\mu). $$

The presence of such a ``deficit'' in the azimuthal angle provides
boundary conditions for quantized fields in the metric (\ref{fmet1}),
which result in Casimir-like distortions of the spectrum of vacuum
fluctuations. The effect of vacuum fluctuations for a massless
spin-$ 1\over 2 $  field has been calculated in \cite{frser} by using
the method of Green's functions. Here we shall briefly reproduce this result
using a different approach, involving the summation of the exact field modes.

   The propagation of a massless spin-$ 1\over 2 $ field in curved
space-time is governed by the Dirac equation
\begin{equation}
i\,\gamma^\mu\,\nabla_\mu\,\psi=0
\label{dirac}
\end{equation}
where the $ \gamma^\mu $ matrices satisfy the anticommutation relations
\begin{equation}
\{\gamma^\mu\, ,\gamma^\nu\}=\,2 \,g^{\mu\nu} .
\label{anticom1}
\end{equation}
The spinor covariant derivative operator is defined as 
$$ \nabla_\mu = \partial_\mu-\Gamma_\mu $$
where  $ \Gamma_\mu $ is the spinor connection. Let us consider a tetrad of
vector field $ e^\mu_{a} $, satisfying the relations 
\begin{equation}
\eta_{ab} =g_{\mu\nu}\,e^\mu_{a}\,e^\nu_{b}, \quad\qquad
e^\mu_{a}\,e^\nu_{b}\,\eta_{ab} = g^{\mu\nu}
\label{tetrad}
\end{equation}
For the sake of further convenience we shall use the Newman-Penrose 
tetrad of null vectors \cite{newp}
\begin{equation}
e^\mu_a=\,\{l^\mu,\, n^\mu,\, m^\mu,\, m^{*\mu}\}
\label{nptetr}
\end{equation}
which satisfy the normalization conditions $\; l_\mu n^\mu= 1 \;$ and
$\; m_{\mu} {m^{*}}^\mu = -1 \; $. The symbol $*$ denotes a complex 
conjugation. It is clear that with the choice of the tetrad (\ref{nptetr}) 
the symmetric constant matrices $ \eta_{ab} $ and $ \eta^{ab} $ have the only
nonvanishing components
$ \;\eta_{12} = \eta_{21} = - \eta_{34} = - \eta_{34} = 1, \; \;\;\;\;
 \eta_{ab}  =  \eta^{ab} $.
Since the ordinary constant Dirac matrices $ \gamma^a $ satisfy the
condition
\begin{equation}
\{\gamma^a,\, \gamma^b\}= 2 \,\eta^{ab},
\end{equation}
then the coordinate-dependent matrices $ \gamma^\mu $ obeying the relation
(\ref{anticom1}) will be defined through the above introduced 
tetrad $ e^\mu_a $ by
\begin{equation}
\gamma^\mu=\,e^\mu_a\, \gamma^a
\label{gammat}
\end{equation}
while the spinor connection $ \Gamma_\mu $ is defined as
\begin{equation}
\Gamma_\mu=\,{1\over 4}\,\gamma^a\,\gamma^b \,e^\nu_{a\,;\,\mu}\,e_{b\,\nu}
\label{spcon}
\end{equation}
where the semicolon denotes the covariant differantation with respect to the
metric $ g^{\mu\nu} $. 
The explicite form of the $\gamma^\mu$ matrices written in terms of the 
null vectors is
\begin{equation}
\gamma^\mu=\sqrt{2}\left(\matrix{0&0&n^\mu&-m^{*\mu}\cr
                0&0&-m^{\mu}&l^\mu\cr
                l^\mu& {m}^{*\mu}&0&0\cr
                m^\mu&n^\mu&0&0\cr}\right)
\label{matrice}
\end{equation}

   Turning to the cosmic string metric (\ref{fmet1}) one may choose the
following null vectors
\begin{eqnarray}
l^\mu =  \frac{1}{\sqrt{2}}\,( 1,\,-1,\,0,\,0 )\,\,\,\,\, &
\qquad
n^\mu & =\frac{1}{\sqrt{2}}\,( 1,\,1,\,0,\,0 )\nonumber\\
&  & \nonumber\\
m^\mu =  {1\over\sqrt{2}}\,( 0,\,0,\,-1,\,-{i\over r} ) &
\qquad
m^{*\mu} & ={1\over\sqrt{2}}\,( 0,\,0,\,-1,\,{i\over r} )
\label{tetr1}
\end{eqnarray}

    Let us now consider the four component spinor 
\begin{equation}
\psi=\left(\matrix{\psi_1\cr
\psi_2\cr \psi_3\cr \psi_4\cr}\right)
\end{equation}
then the Dirac equation (\ref{dirac}) projected onto the null vectors
(\ref{tetr1}) is decomposed into the two independent pairs of equations
$$ D\,\psi_1\,+\,(\delta^*+\,\beta)\,\psi_2=\,0 $$
\begin{equation}
\Delta\,\psi_2\,\,\,+\,(\delta+\,\beta)\,\psi_1=\,0
\label{dirac1}
\end{equation}
and
$$\Delta\,\psi_3\,-\,(\delta^*+\,\beta)\,\psi_4=\,0$$
\begin{equation}
D\,\psi_4\,\,\,-\,(\delta+\,\beta)\,\psi_3=\,0
\label{dirac2}
\end{equation}
where
 $$ \beta = \frac{1}{2} \; {m^*}^{\mu}_{\;\; ; \; \nu} \; m_{\mu} m^{\nu}=
 - \frac{1}{2 \sqrt{2}\, r} $$ 
and
\begin{eqnarray}
D &=&l^\mu\partial_\mu ={1\over{\sqrt{2}}}\,(\partial_t-\partial_z)
\quad\quad 
\Delta = n^\mu\partial_\mu = {1\over{\sqrt{2}}}\,(\partial_t+\partial_z)
\nonumber\\
\delta &=& m^\mu\partial_\mu = -{1\over{\sqrt{2}}}\,
(\partial_r+{i\over r}\,\partial_\varphi)\qquad\qquad\qquad\qquad
\end{eqnarray}
are the directional derivative operators.
Combining the equations (\ref{dirac1}) and (\ref{dirac2}) in an 
appropriate way we obtain the decoupled set of equations
\begin{equation}
\Box_s\psi_s=0
\label{deceq}
\end{equation}
where we have introduced the spin-weighted operator
$$ \Box_s = \partial_t^2 - \partial_z^2 - {1 \over r}\,\partial_r\,
(r\partial_r)
 - {1\over r^2}\,\partial_\varphi^2 - {2is\over r^2}\,\partial_\varphi +
{s^2\over r^2} $$
with the spin-weight $ s = \pm{1\over 2} $ and the spin-weighted modes
$$  \psi_{(s = +{1\over 2})} =\left(\matrix{\psi_2\cr
\psi_4\cr}\right),\qquad\qquad
\psi_{(s = -{1\over 2})} = \left(\matrix{\psi_1\cr
\psi_3\cr}\right) $$
The equations (\ref{deceq}) may be solved by the seperation of variables
and the regular at $\,r=\,0\,$ solutions have the form
\begin{equation}
\psi_s\sim\,e^{-i(\omega t-kz-m\nu\varphi)}
J_{\vert m\nu+s\vert}(p\,r)
\label{modes}
\end{equation}
where $ J(p\,r) $ is a Bessel function and $p^2=\,\omega^2-\,k^2\,$ ,\,\,
$\nu=\,{1/b} $. Since the transformation properties of spinors implies
that a $ 2\, \pi $ rotation of a spinor changes its sign \cite{ryd}, 
the single-valuedness of the solutions (\ref{modes}) requires that
$ m = n - \frac{1}{2} $ , or $ m = n + \frac{1}{2} $  with $ n = 0,\, 
\pm 1\, \pm 2 ... $ . Having in hand these modes one can
easily construct the complete sets of positive- and negative-frequency
solutions $ { u_{smkp} } $ and $ { v_{smkp} } $ of the Dirac equation,
which satisfy the normalization conditions
\begin{equation}  \begin{array}{rr}
\int{\overline u}_{s' m' k' p'}(x)\,\gamma^0\, u_{s\,m\,k\,p}(x)
\sqrt{-g}\, d^3\,x=  
\int{\overline v}_{s' m' k' p'}(x)\,\gamma^0\, v_{s\,m\,k\,p}(x)
\sqrt{-g}\, d^3\,x     & \\ [3mm]
= \delta_{ss'}\,\delta_{mm'}\,\delta(k - k') \,
\frac{\textstyle {\delta(p-p')}}{\textstyle{\sqrt{p\,p'}}}
\end{array}
\end{equation}
where ${\overline u}=\,u^\dagger\,\gamma^0 $ and  
${\overline v}=\,v^\dagger\,\gamma^0 $ are the adjoint spinors.

   The canonical quantization are performed in a usual way \cite{birdav},
by expanding the field operator
$\hat\psi\,(x)$ on the complete sets of positive and negative frequency
modes
\begin{equation}
\hat\psi\,(x)=\,
\sum_{s,m}\int_{-\infty}^\infty dk\int_0^\infty dp\,p\,
[\hat a_{s m k p}\,u_{s m k p}(x)+
\hat b^\dagger_{s m k p}\,v_{s m k p}(x)]
\label{quant}
\end{equation}
where  $\hat a_{s m k p}$ represents annihilation operator for particles and
$\hat b^\dagger_{s m k p}$ is the creation operator for antiparticles.
It should be noted that as the spacetime is flat everywhere outside cosmic strings
one can define a vacuum state by choosing positive frequency
modes with respect to the timelike Killing vector $ \partial/\partial t $
of the flat spacetime. 

   The energy-momentum tensor for spin- $ 1\over 2 $ has the form
\begin{equation}
T_{\mu\nu}(x)={i\over 2}\,[\,\overline{\psi}\,\gamma_{(\mu}\nabla_{\nu )}
\psi-(\nabla_{(\mu}\overline{\psi})\,\gamma_{\nu)}\psi\,]
\label{emt}
\end{equation}
The vacuum expectation value of this quantity
can be evaluated by representing it as a bilinear function of the fields and
performing a renormalization procedure at the coincidence points $x=\,x'.$
Substituting the expansion  (\ref{quant}) into the equation (\ref{emt}) 
we obtain that its vacuum averaged $ < T_{00}\,(x) > $ component are reduced 
to the form
\begin{equation}
< \,T_{00}\,(x)\,>  =\,\lim_{x\rightarrow x'}\,
(\partial_t^2-\,\partial_t\partial_{t'}){U(x,\,x')\over {2\omega}}
\label{ed}
\end{equation}
where the two-points function $ U( x,x') $ is
\begin{eqnarray}
{U(x,x')\over{2\omega}}&=&{\nu\over{8 \pi^2}}
\sum_{m} e^{i\nu m(\varphi-\varphi')}
\int_{0}^\infty dp\,p
\int_{-\infty}^\infty dk\,
{e^{-i\sqrt{k^2+p^2}\,\,\tau 
+\,ik\,\zeta}\over{\sqrt{k^2+p^2}}}\nonumber\\
& &\nonumber\\
&&[J_{\vert m\nu+{1\over 2}\vert}(p\,r)
J_{\vert m\nu+{1\over 2}\vert}(p\,r')+\,
J_{\vert m\nu-{1\over 2}\vert}(p \,r)
J_{\vert m\nu-{1\over 2}\vert}(p\,r')]
\label{bifunction}
\end{eqnarray}
and
$\tau=\,t-t'\,$, $\,\zeta=\,z-z'.$
The integrals over $ k $ and $ p $ in this expression are evaluated
by means of the corresponding formulae given in \cite{gr}. As a result
we have
\begin{equation}
{U(x,x')\over{2\omega}} ={\nu\over{8\pi^2\,r\,r'}}
{1\over\sqrt{u^2-1}}\sum_{n=-\infty}^\infty 
e^{-i\nu\,(n- {1\over 2})(\varphi-\varphi')}\,
\left( \,\xi^{-\vert\nu (n-{1\over 2})+{1\over 2}\vert}+
\xi^{-\vert\nu (n-{1\over 2}) - {1\over 2}\vert}\,\right)
\label{bisimple}
\end{equation}
where
$$
u={r^2+\,r'^2+\zeta^2-\tau^2\over{2\,r r'}},\qquad\qquad\xi=
{u+\sqrt{u^2-1}} $$
and we have taken  $ m= n-{1\over 2} $.
The renormalization of the function $ U(x,\,x') $ is achieved by 
substracting from this expression its pure Minkowskian value, i.e at
$\nu=\,1 $. Since the equation (\ref{ed}) involves only the derivatives
over $t$ and $t'$ one can put $\varphi=\,\varphi',$  $z=\,z'$ and $r=\,r'$.
Then the evaluation of the sum in  (\ref{bisimple}) is significantly
simplified and we have
\begin{equation}
\sum_{n=-\infty}^\infty\, \left(\xi^{-\vert\nu (n-{1\over 2})+{1\over 2}\vert}+
\xi^{-\vert\nu (n- {1\over 2}) -{1\over 2}\vert} \right)= 2\,\,
{\xi^{-{\nu\over 2}}\,(\xi^{-{1\over 2}}+\xi^{1\over 2})\over{1-
\xi^{-\nu}}}
\label{bifinal}
\end{equation}
The further calculation of the renormalized quantity $ \, U(x,\,x') \,$
becomes straightforward and substituting the result into the equation 
(\ref{ed}) we arrive at the following expression for the energy 
density of vacuum fluctuations
\begin{equation}
< T_{00}(x)>  =-{\hbar\over 2880}
{1\over\pi^2\,r^4}\,\,(\nu^2\,-1)\,(7\nu^2+\,17)
\label{exed}
\end{equation}
This result is in agreement with that of given in \cite{frser}. The other 
components of $\; < T_{\mu\nu}\,(x) > \; $  can be found by using 
the symmetry properties of the space-time  (\ref{met1}) along with 
the conditions $\,\,T^\mu_\mu=0\,\,$ and 
$\,\,T^{\mu\nu}_{\,\,\,\,\,\,;\,\nu}=\,0.$

\section{ Multiple cosmic strings }

  If one makes a transformation of the radial coordinate $r\rightarrow
{r_i^b/b}$, then the metric (\ref{met1}) can be transformed into the
the following form
\begin{equation}
ds^2=\,dt^2-\,dz^2-\,e^{-2 \Lambda(x,y)}(dx^2+dy^2)
\label{met2}
\end{equation}
where
\begin{equation}
\Lambda(x,y)=\sum_{i=1}^N 4 \,\mu_i\ln r_i\quad\quad
r_i=[(x-\alpha_i)^2+\,(y-\beta_i)^2]^{1/2}
\end{equation}
It turns out that this metric is the exact solution of the Einstein 
equations \cite{let}, describing the space-time around  $N$ parallel static
cosmic strings, passing through the points $x_i=\,(\alpha_i,\,\beta_i)$.
The corresponding tetrad of null vectors for the metric (\ref{met2})
can be chosen as
\begin{eqnarray}
l^\mu =  \frac{1}{\sqrt{2}}\,(1,\,-1,\,0,\,0)\qquad\, &
\qquad
n^\mu & =\frac{1}{\sqrt{2}}\,(1,\,1,\,0,\,0)\nonumber\\
&  & \nonumber\\
m^\mu =  {1\over\sqrt{2}}\,e^\Lambda\,(0,\,0,\,-1,\,-i) &
\qquad
m^{*\mu} & =-{1\over\sqrt{2}}\,e^\Lambda\,(0,\,0,\,1 \,-i)
\label{tetr2}
\end{eqnarray}
Using these vectors in the equations (\ref{gammat}) and (\ref{spcon})
we find that  
\begin{equation}
\gamma^0(x)=\, \gamma^0, \;\; \gamma^3(x)= \gamma^3, \;\;
\gamma^1(x)= e^\Lambda\gamma^1,\;\; \gamma^2(x)= e^\Lambda\gamma^2,
\label{gamtetr}
\end{equation}
where $\gamma^0, \;\; \gamma^1, \;\;\gamma^2, \;\;\gamma^3 \;  $
are ordinary constant Dirac matrices, and
the only non-vanishing components of the spinor connection $\Gamma_\mu$
are
\begin{equation}
\Gamma_1=- \,{1\over 2}\,\gamma^1\gamma^2\,e^\Lambda\,\partial_y\Lambda
\qquad\Gamma_2= - \,{1\over 2}\,\gamma^2\gamma^1\,e^\Lambda\,\partial_x\Lambda
\label{spcon1}
\end{equation}
Now we shall evaluate the vacuum expectation value of the energy-momentum
tensor (\ref{emt}) in the metric (\ref{met2}).
We start with the expression 
\begin{equation}
<T_{\mu\nu}(x)> =\,-{\hbar\over 4}\lim_{x\rightarrow\,x'} 
{\rm Tr}\,[\gamma_\mu(\nabla_\nu-\nabla_{\nu'})+\,\gamma_\nu(\nabla_\mu
-\nabla_{\mu'})]S_F(x,\,x')
\label{emt1}
\end{equation}
where
\begin{equation}
S_F(x,\,x')=\,-i<  0\vert\,\,T\left(\bar{\psi}(x')\,{\psi}(x)\right)
\vert 0>
\label{fp}
\end{equation}
is the Feynman propagator which obeys the equation
\begin{equation}
i\gamma^\mu(\partial_\mu-\Gamma_\mu)S_F(x,x')=\,
{1\over{\sqrt{-g}}}\,\,{\delta^4(x-x')}
\end{equation}
The substitution of the relations (\ref{gamtetr}) and  (\ref{spcon1})
into this equation
enables us to cast it in the form 
\begin{equation}
i\gamma^a \,\partial_a \,S_F(x,x')=\,{\delta^4(x-x')}+\,V\,S_F(x,x')
\label{diracexp}
\end{equation}
where $\gamma^a $ once again denotes the flat space-time Dirac matrices,
\begin{equation}
V= i\,[(1-e^{-2\Lambda})\gamma^A\partial_A+\,
(1-e^{-\Lambda})\gamma^\alpha \partial_\alpha+\,
{1\over 2}e^{-\Lambda}\gamma^\alpha \partial_\alpha \Lambda]
\label{potential}
\end{equation}
and the index $A$ takes the values $(0,3)\equiv\,(t,z),$ while
$\alpha=(1,2)\equiv\,(x,y).$ 
In order to construct the solution of equation (\ref{diracexp})
we use a perturbative approach. For this purpose, we assume that
the linear mass densities of the cosmic strings are sufficiently small,
$(\mu_i << 1),$ which is indeed the case for realistic cosmic strings
($\mu_i \approx 10^{-6}$)
Then the metric (\ref{met2}) can be expanded in powers of $\mu_i,$ 
about a fixed flat background, and the potential (\ref{potential}) 
can be considered as a small perturbing term in the equation (\ref{diracexp}). This
approximation allows us to write the solution of equation (\ref{diracexp})
as perturbation series
\begin{equation}
S_F(x,x')=\,S_F^{(0)}+\,S_F^{(0)} V S_F^{(0)}+\,S_F^{(0)} V S_F^{(0)} V S_F^{(0)}
+\,\cdots
\label{expansion}
\end{equation}
The zeroth order free Feynman propagator is defined as 
\begin{equation}
S_F^{(0)}(x,x')=\,\int{d^4k\over{(2\pi)^4}}\,{\gamma k\over k^2}\,
e^{-ik(x-x')}
\label{freep}
\end{equation}
where  $ \gamma k = \gamma^a k_a $. The successive terms in the
expansion (\ref{expansion}) correspond to the higher order in $\mu_i$
contributions to the free propagator. Expanding the equation
(\ref{potential}) in powers of $ \mu_i $ and taking it into account
in (\ref{expansion}) we obtain that the first order corrections
to the free Feynman propagator are given by
$$
S_F^{(1)}(x,x')=\,{1\over 64\pi^6}
\int{d^4k\,\delta(k_0)\,\delta(k_3)}\,\Lambda(k)\,e^{-ikx'}
\int{d^4q\over q^2\,(q-k)^2}\, \gamma q$$
\begin{equation}
\qquad\qquad[\,\gamma^A q_A+\,\gamma q-{1\over 2} \gamma k\,]\,
\gamma (q- k)\,e^{-iq(x-x')}
\label{fp1}
\end{equation}

   In order to evaluate the finite vacuum expectation value of the
energy-momentum tensor(\ref{emt1})  one needs to regularize
it, for which we shall use the dimensional regularization procedure 
\cite{thof}. The latter gives to the the vanishing value for the 
zeroth order propagator \cite{birdav}, however, substituting the first-order
propagator (\ref{fp1}) into the expression (\ref{emt1}),
for its $<T^{(1)}_{00}(x)>$ component at the coincidence limit
$x\rightarrow x'$  we find
$$<T^{(1)}_{00}(x)>  = {i\hbar\over 64\pi^6}
{\rm Tr}\,\int\,d^4k\,\delta(k_0)\,\delta(k_3)\,\Lambda(k)\,e^{-ikx'}
\int\,{d^4q\over q^2(q-k)^2}$$
\begin{equation}
\qquad\qquad\qquad\gamma^0 q_0\,\gamma q\,
[\,\gamma^A q_A+\,\gamma q-{1\over 2}\gamma k\,]\,\gamma (q-k)
\end{equation}
In the framework of the dimensional regularization procedure the
integral over $ q $ is calculated, by performing an analytical
continuation to $d$ dimensional space and then the result is expanded
about $d=\,4-\epsilon,$ ($\epsilon\rightarrow 0$) \cite{ryd}. 
Having done all these we obtain 
\begin{equation}
<T^{(1)}_{00}(x)>  = -\;{\hbar\over 240\pi^3}\Gamma(-1+{\epsilon\over 2})
\sum_{i=1}^N\,\mu_i\,\int d^2k\,(k^2)^{1-{\epsilon\over 2}}\,e^{-ikr_i}
\label{enerjid}
\end{equation}
In obtaining of this expression we have used the explicit form for the
Fourier component of the function $\, \Lambda(x,y) \,$
\begin{equation}
\Lambda(k)=\,8\pi\sum_{i=1}^N\,
\mu_i\,{e^{ik\,x_i}\over k^2}
\label{fourier}
\end{equation}
It is important to stress that the expression (\ref{enerjid}) is finite,
as the divergent terms are exactly cancelled by one another. Indeed, using the
integral \cite{ali3}
\begin{equation}
\int_0^\infty d^dk(k^2)^\nu e^{-ikR}=
\pi^{d/2}\,{2^{2\nu+d}\over{R^{2\nu+d}}}\,
{\Gamma(\nu+{d/2})\over{\Gamma(-\nu)}}
\label{mainint}
\end{equation}
in Eq.(\ref{enerjid}), in which
$d=2-\epsilon,\,$ and $ \nu=\,1-{\epsilon\over 2} $, we finally
obtain the following result
\begin{equation}
<T^{(1)}_{00}(x)>  =-{\hbar\over 15\pi^2}\sum_{i=1}^N
{\mu_i\over{r_i^4}}
\label{enerji1}
\end{equation}
It is seen from this expression that at first order in $\mu_i$,
the contributions to the energy density of vacuum fluctuations of
spin-$1\over 2$ field around $N$ parallel static cosmic strings are 
linearly summed. One can easily see that in the case of a single 
cosmic string the expression (\ref{enerji1}) coincides with  
(\ref{exed}) taken at $\mu <<1$.
\begin{center}
\section{Vacuum force between two parallel cosmic strings}
\end{center}

   Let us now proceed to the second-order metric contributions to the vacuum
expectation value of the energy-momentum tensor (\ref{emt1}). In analogy
with the case of electromagnetic fluctuations \cite{ali3} we shall show 
that the energy density of vacuum fluctuations of a massless spinor 
field involves a term which depends upon the seperation distance 
between cosmic strings, therefore, produces an attractive force 
between the strings.

Using the expressions (\ref{potential}) and(\ref{freep}) in the expansion
(\ref{expansion}) we find that the second-order metric corrections
to the free Feynman propagator have the form
$$
S_F^{(2)}(x,x)=\,{1\over 256\pi^8}\,
\int d^4 k\,\delta(k_0)\delta(k_3)\,\Lambda(k)e^{-ik\,x'}
\int d^4 p\,\delta(p_0)\delta(p_3)\,\Lambda(p)e^{-ip\,x'}$$
$$
\int\,{d^4 q\over q^2}\,{\gamma q\,\,(\gamma^A q_A+\,\gamma q-{1\over 2} 
\gamma k)\,\,\gamma(q-k)\over (q-k)^2\,\,(q-k-p)^2}\qquad\qquad
$$
\begin{equation}
\qquad\qquad
[\gamma^B q_B\,+\gamma(q-k-p)\,+{1\over 2}\gamma p]\,\,
\gamma(q-k-p)\,e^{-iq\,(x-x')}
\label{fp2}
\end{equation}
Substituting this expression into the $ < T_{00} > $ component
of Eq. (\ref{emt1}) at the
coincidence limit  $x=\,x'$, we calculate the traces of $ \gamma $ matrices
using the well-known theorems  \cite{ryd}, then taking into account 
the relation (\ref{fourier}) we arrive at the expression
$$
<T^{(2)}_{00}(x)>  ={i\hbar\over 4\pi^6}\sum_{i=1}^N\mu_i
\sum_{j=1}^N\mu_j 
\int\,{d^4 k\over k^2}\,\delta(k_0)\,\delta(k_3)\,e^{-ik\,(x-x_i)}
\qquad\qquad\qquad$$
\begin{equation}
\qquad\int\,{d^4 p\over p^2}\,\delta(p_0)\,\delta(p_3)\,e^{-ip\,(x-x_j)}
\int\,{d^4 q\over q^2}\,{q_0^2\over (q-k)^2(q-k-p)^2}\,\,{\cal N}(q,\,k,\,p)
\label{ed2}
\end{equation}
where 
$${\cal N}(q, k, p)=4(q_A\,q^A)\big[4\,q_A\,q^A\,+\,(q-k)^2+\,(k+p)^2  - kp  
\big] +\,2(q^2-\,2\,kq)\,(kq-\,2 q^2)$$
$$\qquad
-\,k^2\,(2\,q^2+\,2\,kq-\,p^2-\,2\,k^2)
-\,(kp)\,(q^2-\,3\,k^2+\,2\,kq)+\,2\,(pq)\,(q-k)\,(3q-k)$$
We note that in this expression we are interested only in contributions to the vacuum
energy density which are proportional to the products of the linear 
mass densities of different cosmic strings. It is clear that the latter 
describes the energy density of vacuum interaction between the strings. 
As for the remaining contributions,
they form higher order corrections to the vacuum energy density given by
(\ref{enerji1}). However first we need to regularize the
expression (\ref{ed2}), for which we again use the dimensional regularization
procedure. We evaluate the integral over $q$ by performing an analytical 
continuation to $ d=4-\epsilon $ dimensions using the scheme, described in 
\cite{ryd}. After some straightforward algebra, we arrive at 
the following result 
\vspace{4mm}
$$
<T^{(2)}_{00}(x)>=-{\hbar\over 4\pi^4}\sum_{i=1^N}\mu_i
\sum_{j=1}^N\mu_j 
\int\,{d^2 k\over k^2}\,e^{-ik\,(x-x_i)}
\int\,{d^2 p\over p^2}\,e^{-ip\,(x-x_j)}\qquad$$
\vspace{4mm}
$$\qquad\qquad
\int_0^1\,dz_1 \,\int_0^{1-z_1}\,dz_2 \,\big[40\,\Gamma(-2+{\epsilon\over 2})\,
B^{2-{\epsilon\over 2}}+\,C_1\,\Gamma(-1+{\epsilon\over 2})
\,B^{1-{\epsilon\over 2}}$$
\vspace{4mm}
\begin{equation}
+\,C_2\,\Gamma(\,{\epsilon\over 2}\,)\,B^{-{\epsilon\over 2}}\big]\qquad\qquad\qquad\qquad
\qquad\qquad\qquad\qquad
\label{enerji2}
\end{equation}
where we have introduced
$$ \quad C_1= \,-12Q\,(Q-\,k-\,p)+4\,(k^2+\,p^2+\,{1\over 8}kp\,)
\quad\qquad\qquad\qquad $$
$$C_2= \,Q^2\,(-2\,Q^2+\,5\,kQ-\,k^2-{1\over 2}kp+\,3pQ\,)
\qquad\qquad\qquad\qquad$$
$$\qquad\qquad
-\,kQ\,(\,k^2+\,kp+\,2kQ+\,4pQ\,)\,+\,{1\over 2}k^2\,
(\,p^2+\,2k^2+\,3kp+\,2pQ\,)\, $$
and
$$
Q_\alpha=\,k_\alpha\,z_1+\,(k_\alpha\,+\,p_\alpha)\,z_2
\qquad\qquad\qquad\qquad\qquad\qquad\qquad $$

$$ B=\,k^2\,z_1(1-z_1)+\,(k+p)^2\,z_2\,(1-z_2)-\,2k\,(k+p)\,z_1z_2 $$

   The energy of vacuum fluctuations per unit length of the strings may be
evaluated by means of the formula
\begin{equation}
E=\int\,d^2 x \sqrt{-g} < T_{00}(x) >
\label{int0}
\end{equation}
which at the second order metric perturbations takes the following form
\begin{equation}
E=\,\int\,d x d y \Big[\,<T_{00}^{(2)}\,(x)>\,\,-\,\,2\Lambda(x)
<T_{00}^{(1)}\,(x)>\Big]
\label{int}
\end{equation}
For the sake of certainty let us consider two parallel cosmic
strings. Using the relations (\ref{enerjid}) and (\ref{enerji2}) in
the equation (\ref{int}) we first carry out the integration over $ x $ 
and $ y $, then the calculations of remaining integrals
become straightforward and keeping only the terms involving the product of
the linear mass densities of the cosmic strings we find
\begin{equation}
E_{int}=\,-{\hbar\over 15\pi^2}\,\mu_1\mu_2\,\Gamma(\,{\epsilon\over 2}\,)
\int\,d^2 k\,(k^2)^{-{\epsilon\over 2}}\,e^{-ika}.
\label{int1}
\end{equation}
It should be stressed that this quantity is finite as the involved divergent
at $ \epsilon\rightarrow 0 $ terms are compensated by one another.
Indeed taking the double integration over $ k $ using the integral
(\ref{mainint}) we obtain the expression
\begin{equation}
E_{int}=\,-{4\hbar\over 15\pi}\,{\mu_1\mu_2\over a^2}
\end{equation}
where $ a $ is the seperation distance between the cosmic strings.
It is clear that the presence of this energy, will produce an
attractive force per unit length of the strings given by
\begin{equation}
F=\,-{8\hbar\over 15\pi}\,{\mu_1\mu_2\over a^3}
\label{force}
\end{equation}

    As we have already mentioned above the static and straight-line 
cosmic strings do not exert any local gravitational force on surrounding 
matter. Here we have shown that the propagation of a massless quantized spinor
field in the spacetime of more than one cosmic string induces a force of
attraction (\ref{force}) between two cosmic strings, 
which falls of as the third power of the separation distance.
The reason for this is the restriction of the modes of quantized field
by the multiconical structure of the spacetime around the cosmic strings.
Unlike the case of a single cosmic string, it is difficult to construct the
exact modes of the field equations in the metric of multiple
cosmic strings, so we have used a perturbative approach along with the
dimensional regularization procedure. We note that the expression
(\ref{force}) coincides with the corresponding result for a massless
scalar field \cite{gal}, while 2 times smaller than the result for a
massless vector field \cite{ali3}.

   It is important to stress that the above result is obtained within the
 one-loop approximation and therefore holds provided that the separation
distance between the cosmic strings is much greater than the typical
thicknesses of their cores.

\end{document}